\documentclass[iop]{emulateapj}
\usepackage{amsmath,amstext}
\usepackage[T1]{fontenc}
\usepackage{apjfonts} 
\usepackage{booktabs}
\usepackage{hyperref}

\newcommand\ionn[2]{#1$\;${\scriptsize\rmfamily{#2}}\relax}

\shorttitle{Primordial Helium Abundance}
\shortauthors{Valerdi et al.}

\begin{document}

\title{Determination of the primordial helium abundance based on NGC~346 an \ion{H}{2} region of the Small Magellanic Cloud}

\author{Mabel Valerdi, Antonio Peimbert, Manuel Peimbert, \& Andr\'es Sixtos}
\affiliation{Instituto de Astronom\'ia, Universidad Nacional Aut\'onoma de M\'exico, Apdo. Postal 70-264 Ciudad Universitaria, M\'exico}
\email{mvalerdi@astro.unam.mx}

\begin{abstract}
To place meaningful constraints on Big Bang Nucleosynthesis models, the primordial helium abundance determination is crucial. Low-metallicity \ion{H}{2} regions have been used to estimate it since their statistical uncertainties are relatively small. We present a new determination of the primordial helium abundance, based on long slit spectra of the \ion{H}{2} region NGC~346 in the small Magellanic cloud. We obtained spectra using three $409'' \times 0.51''$ slits divided in 97 subsets. They cover the range $\lambda\lambda3600-7400$ of the electromagnetic spectrum. We used {\texttt{PyNeb}} and standard reduction procedures to determine the physical conditions and chemical composition. We found that for NGC~346: $X=0.7465$, $Y=0.2505$ and $Z=0.0030$. By assuming $\Delta Y / \Delta O = 3.3\pm0.7$ we found that the primordial helium abundance is $Y_{\rm P}= 0.2451 \pm 0.0026$ (1$\sigma$). Our $Y_{\rm P}$ value is in agreement with the value of neutrino families, $N_{\nu}$, and with the neutron half-life time, $\tau_{n}$, obtained in the laboratory.
\end{abstract}


\section{Introduction}

A highly accurate ($\lesssim 1\%$) primordial helium abundance ($Y_{\rm P}$) determination plays an important role in understanding the Universe. In particular, it is extremely important to constrain: Big Bang Nucleosynthesis models, elementary particle physics and the study of galactic chemical evolution.

To obtain a value for $Y_{\rm P}$ using \ion{H}{2} regions, it is necessary to use the relation between the helium mass fraction ($Y$) and the heavy elements mass fraction ($Z$). The relation is then extrapolated to zero metallicity to estimate the primordial mass fraction of helium. The first determination of $Y_{\rm P}$ based on observations of \ion{H}{2} regions was carried out by \citet{Peimbert1974}. Since then, many estimations have been done using this technique \citep[e.g., ][]{Izotov2014, Aver2015}. This method has been modified to use the O abundance instead of $Z$, because measuring other chemical elements becomes impractical. Although some groups seek alternatives to O, like using N or S \citep[e.g.][]{Pagel1992, Vital2018}.

In recent years, careful studies with high quality determinations of $Y_{\rm P}$ have been done by: \citet{Izotov2014}, \citet{Aver2015}, \citet{Peimbert2016}, and \citet{Vital2018}. To diminish the uncertainties most determinations use many objects; but large samples only diminish the contribution of statistical errors, while systematic errors are not diminished; in fact, if one is not careful, systematic errors can be more significant for large samples, since low quality objects/observations are often included \citep[e.g., ][]{Peimbert2000, Peimbert2007, Ferland2010}.

Low-metallicity \ion{H}{2} regions have permitted to  estimate the primordial helium abundance, whose statistical uncertainties are very small ($\lesssim 1\%$) \citep[e.g.][]{Izotov2014}; nevertheless, due to the numerous systematic uncertainties, obtaining better than 1\% precision for individual objects remains a challenge. 

The Small Magellanic Cloud (SMC) hosts the region NGC~346, this is the best \ion{H}{2} region to determine $Y_{\rm P}$. This is the brightest and largest \ion{H}{2} region of the SMC. Its H$\alpha$ intensity places it on the boundary between normal and giant extragalactic \ion{H}{2} regions. In addition, its proximity, at a distance of $61\pm1$ Kpc \citep{Hilditch2005}, makes it possible to resolve individual stars and to avoid their contributions to the observed nebular spectra. We also note that NGC~346 hosts the largest sample of O-type stars in the entire SMC \citep{Massey1989}.

NGC~346 possesses some definite advantages over other \ion{H}{2} regions: as it is a nearby object, the underlying absorption correction (due to the stellar absorption of the H and He lines) can be reduced significantly, at least by an order of magnitude; another advantage of its proximity is that it can be observed in different lines of sight, this has been used to rule out the presence of a significant helium ionization correction factor \citep[ICF; e.g.][]{Vilchez1989, Mathis1991, Peimbert1992, Viegas2000, Zhang2003}; another important characteristic is that the electronic temperature is smaller than in the less metallic \ion{H}{2} regions, this reduces the effect of collisional excitation of the ground level of H$^0$; its metallicity ($Z\sim0.003$; $\sim20$\% of solar) requires a small extrapolation to derive $Y_{\rm P}$. However, it has the disadvantage that the correction for its chemical evolution, is larger than in many other objects used for the determination of $Y_{\rm P}$ \citep{Peimbert2000}. Overall there is no \ion{H}{2} region that is simultaneously at least as close, as large, and with a metallicity as low as NGC~346.

In this paper, we present a very precise primordial helium abundance determination. For this we use a single \ion{H}{2} region, NGC~346, which is probably the best object for this determination, for the reasons mentioned before, see the comment by Steigman in \citet{Ferland2010}. Additionally, we include a detailed study of the sources of error involved in this determination, along with a justification of using a single \ion{H}{2} region to determine $Y_{\rm P}$.

The paper is organized as follows: Section \ref{sec:observations} presents the characteristics of the observations and the data reduction. The line intensity corrections due to extinction and underlying absorption are presented in Section \ref{sec:reddening}. Section \ref{sec:physical} shows the determination of physical conditions, using both recombination lines (RLs) and collisional excitation lines (CELs). Sections \ref{sec:abundances} and \ref{sec:total-abun} show the determination of chemical abundances, the determination of ionic abundances and of total abundances respectively. Finally, the determination of the primordial helium abundance, and the conclusions are shown in sections \ref{sec:prim-helium} and \ref{sec:conclusions}.

\section{Spectroscopic Observations}
\label{sec:observations}

We obtained long-slit spectra with the Focal Reducer Low Dispersion Spectrograph FORS1, the night of 2002 September 10 at the Very Large Telescope Facility (VLT), located at Melipal, Chile for the \ion{H}{2} region NGC~346 ($\alpha=00^{h}59^{m}05^{s}.0$ $\delta=-72^{\circ}10'38''$). We used three grism configurations: GRIS-600B+12, GRIS-600R+14 with the filter GG435, and GRIS-300V with the filter GG375. 

\begin{table}[htb]
\centering
\caption{Observation Settings. \label{tab:obs}}
\begin{tabular}{lcccc}
\hline
\hline
Grism & Filter & $\lambda$ & Resolution & Exp. time\\
  & & (\AA) & ($\lambda$/$\Delta\lambda$) & (s)\\ \hline
600B+12 & $-$ & 3560 - 5974 & 1300 & $720 \times 3$ \\
600R+14 & GG435 & 5330 - 7485 & 1700 & $600 \times 3$\\
300V & GG375 & 3440 - 7600 & 700 & $120 \times 3$ \\
\hline
\end{tabular}
\end{table}

Table \ref{tab:obs} shows the resolution and wavelength coverage for the emission lines observed with each grism. We chose this resolution as a compromise: seeking high resolution without breaking the spectra into too many segments.

\begin{figure}[htb]
\begin{center}
\includegraphics[width=8.6cm]{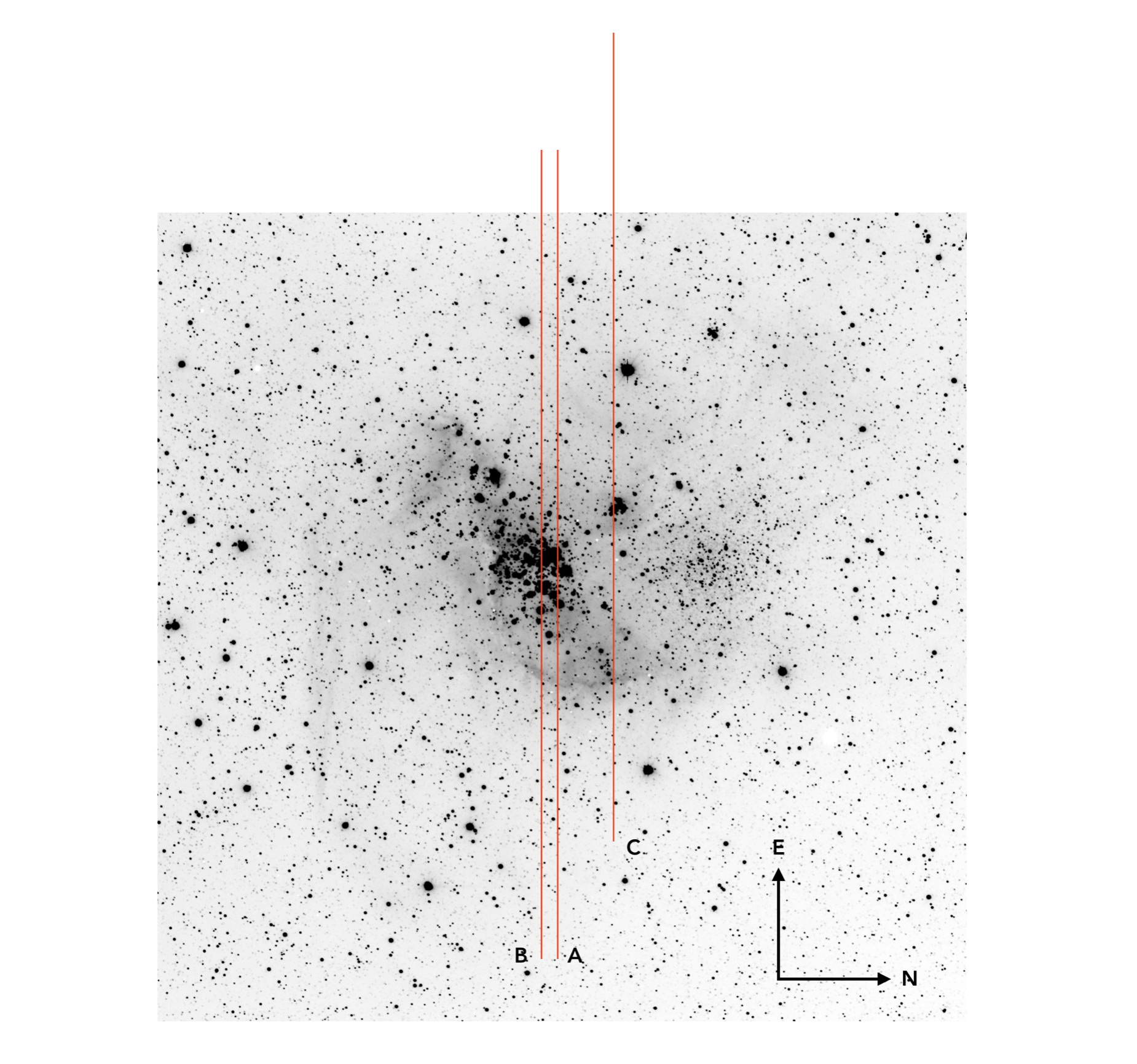}
\caption{VLT images of NGC~346 in the SMC. This object is located at $\alpha=00^{h}59^{m}05^{s}.0$, $\delta=-72^{\circ}10'38''$.}
\label{fig:slits}
\end{center}
\end{figure}

The slit width and length were $0''.51$ and $410''$ respectively. We have three different positions of the slit in the nebula: from the first position A (slit position $\alpha=00^{h}59^{m}06^{s}$ $\delta=-72^{\circ}10'29.3''$), we defined 32 extraction windows; from the second position B (slit position $\alpha=00^{h}59^{m}06^{s}$ $\delta=-72^{\circ}10'37.3''$), we defined 31 extraction windows; and from the last position C (slit position $\alpha=00^{h}59^{m}19^{s}$ $\delta=-72^{\circ}10'0.7''$), we defined 34 extraction windows. The limits for each window were selected to avoid regions with qualitatively different spectra (e.g. stars or clouds) or to break long regions into more homogeneous bits. To keep the same observed region within the slit, we used the LACD (linear atmospheric dispersion corrector) regardless of the air mass value.

We used {\texttt{IRAF}} to reduce the spectra, following the standard procedure; bias, dark and flat correction, wavelength calibration, flux calibration, spectra extraction. We use the low-resolution data to tie the calibration of the blue and red spectra by calibrating the brightest (blue and red) lines to the low-resolution spectra. For the flux calibration we used the standard stars LTT 2415, LTT 7389, LTT 7987, and EG 21 by \citet{Hamuy1992,Hamuy1994}. We did an analysis of the full sample, for each spectrum, in the three slit positions A, B, and C.

\begin{figure}[htb]
\begin{center}
\includegraphics[width=8.6cm]{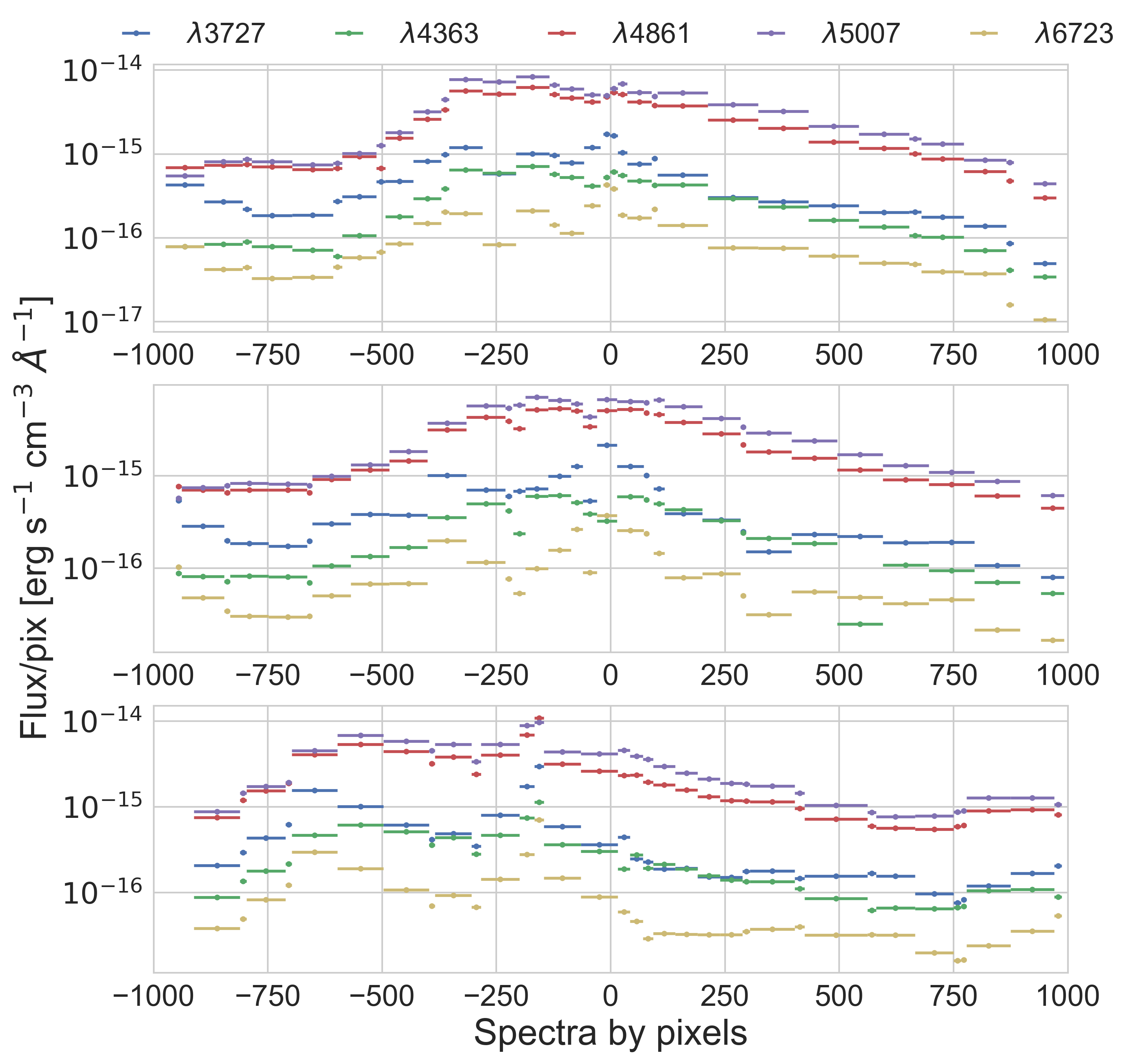}
\caption{The plot shows the intensity of the five emission lines ($\lambda3727$, $\lambda4363$, $\lambda4861$, $\lambda5007$, and $\lambda6717+31$) in the 97 windows for each slit A, B and C (up, middle and down respectively). On the x axis, the bar corresponds to the number of pixels that each window contains. The slits are centered in pixel zero.}
\label{fig:lines}
\end{center}
\end{figure}

\begin{figure}[htb]
\begin{center}
\includegraphics[width=8.6cm]{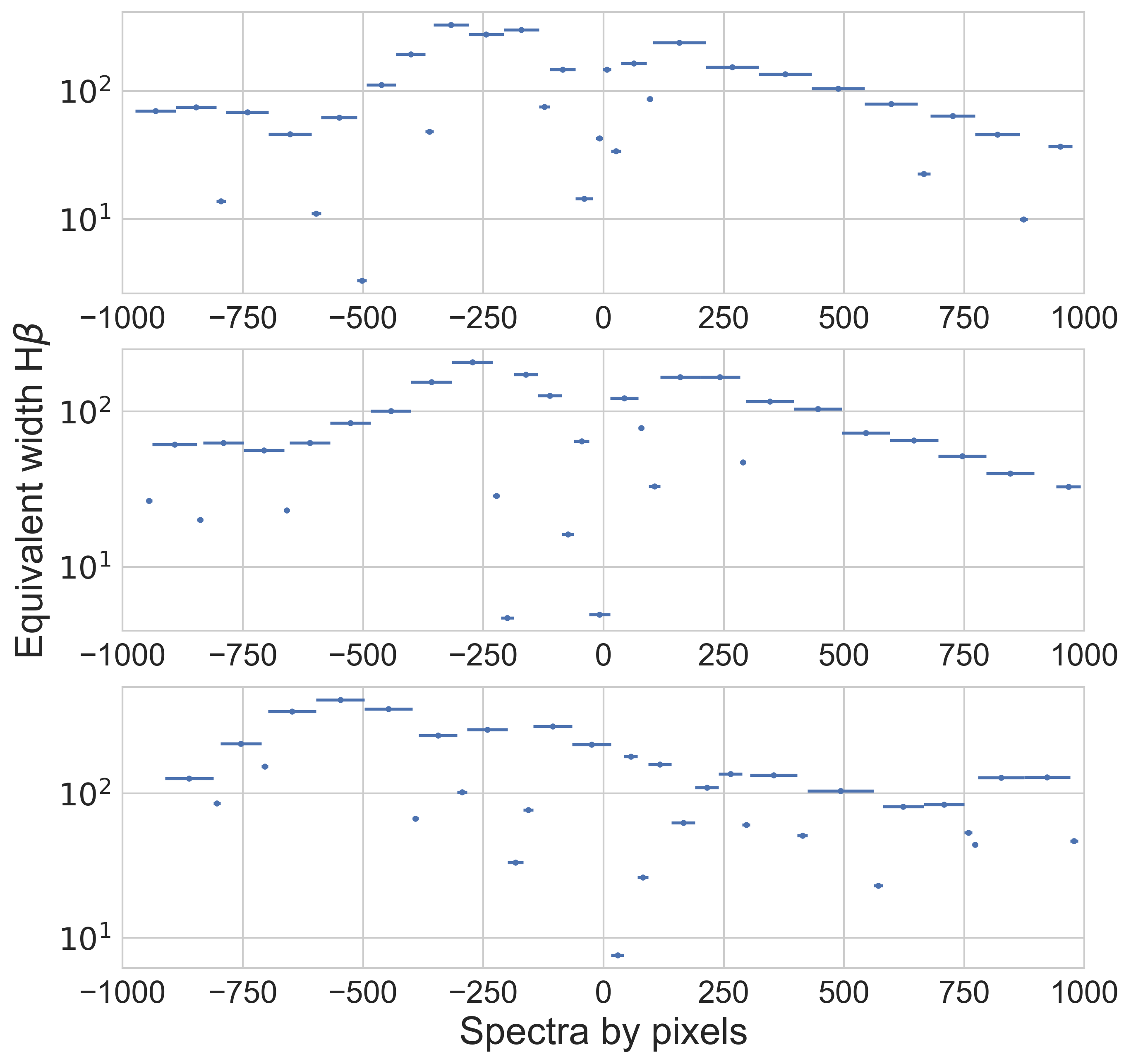}
\caption{The plot shows the EW(H$\beta$) for each window in three slit positions. Section A, B and C (up, middle and down respectively). On the x axis, the bar corresponds to the number of pixels that each window contains. The slits are centered in pixel zero.}
\label{fig:eqw}
\end{center}
\end{figure}

We did a rough analysis of each of our 97 windows; in it we measured H$\beta$ (as a measure of the overall intensity), [\ion{O}{2}] $\lambda$3727 (as a measure of O$^{+}$), [\ion{O}{3}] $\lambda$5007 (as a measure of O$^{++}$), [\ion{O}{3}] $\lambda$4363 (as a measure of the quality of the temperature), and [\ion{S}{2}] $\lambda 6717+31$ (as a measure of the quality of the density); see Figure \ref{fig:lines}. We also present the equivalent width of H$\beta$ in emission (to estimate the influence of the underlying absorption); see Figure \ref{fig:eqw}. In a future paper, we will present the line intensities for each of the 97 windows.

We want to select the windows that cover the brightest parts while avoiding regions where there is too much stellar continuum (i.e. we remove the brightest stars to avoid the effect of stellar underlying absorption, see Figure \ref{fig:slits}).

To minimize the errors of the emission lines we define a new region, Region I, summing the windows with best data. We selected which windows to include in Region I using 3 criteria: they needed to be bright, to have a high equivalent width in emission, and latter we will see that there was some Wolf -Rayet \ion{He}{2} emission that we tried to diminish. We find that windows with a high equivalent width in emission are brighter than $I({\rm H}\beta)=2\times10^{-15}$ erg s$^{-1}$ cm$^{-3}$ \AA$^{-1}$ per pixel.

\begin{figure}[htb]
\begin{center}
\includegraphics[width=8.6cm]{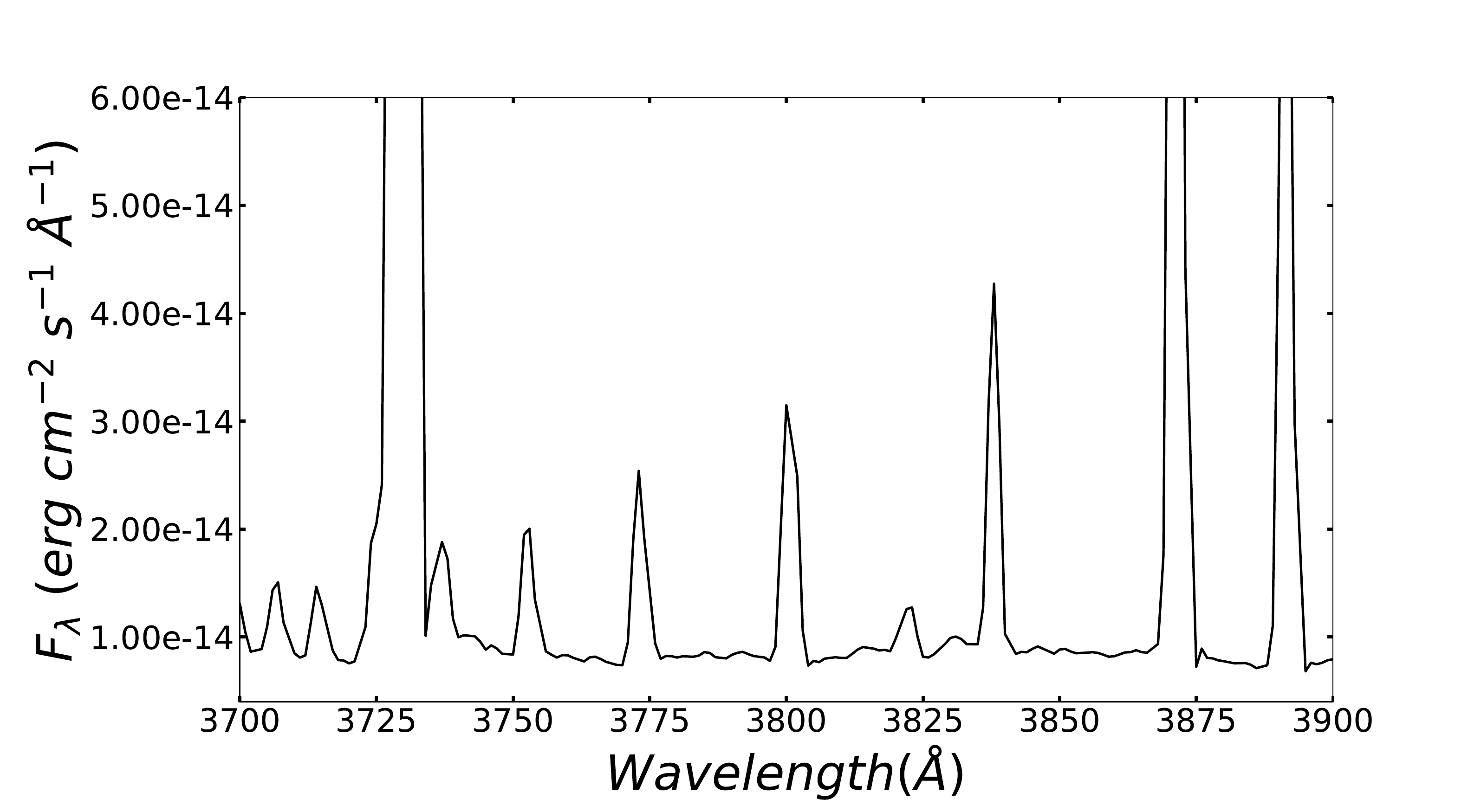}
\caption{Plot of a fraction of the observed wavelength blue range for the high-resolution spectrum of the chosen extractions.} 
\label{fig:blue}
\end{center}
\end{figure}

\begin{figure}[htb]
\begin{center}
\includegraphics[width=8.6cm]{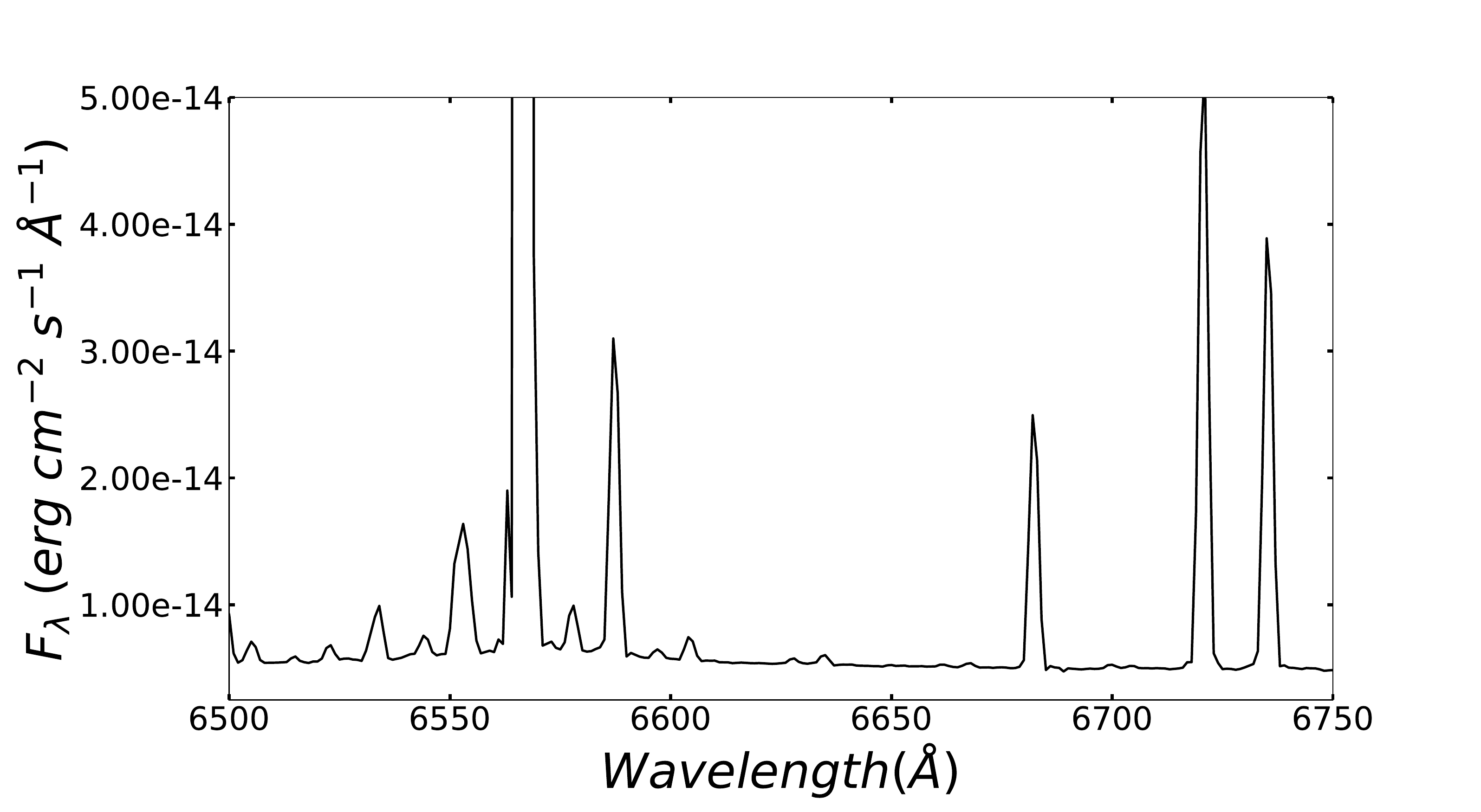}
\caption{Plot of a fraction of the observed wavelength red range for high-resolution spectrum of the chosen extractions.} 
\label{fig:red}
\end{center}
\end{figure}

Finally, we are left with regions that have ${\rm Eqw}({\rm H}\beta)>$ 150, and that do not show Wolf-Rayet contamination in \ion{He}{2} $\lambda$4686. Region I was lastly composed of 20 windows (A10, A12, A13, A14, A21, A23, A24, B10, B11, B14, B22, B23, C3, C5, C6, C7, C9, C11, C14, and C15). In Figures \ref{fig:blue}, and \ref{fig:red}, we show the blue and red spectra of Region I.

\section{Line intensities and reddening correction}
\label{sec:reddening}

Emission line flux measurements were made with the {\texttt{SPLOT}} routine of the {\texttt{IRAF}} package. These were measured by integrating all the flux in the line between two given limits and by subtracting the local continuum estimated by eye. Partially blended lines were deblended with two or three Gaussian profiles to measure the individual line fluxes and we forced the lines to have the same widths.

We derived the reddening by fitting the observed flux ratios of the brightest Balmer lines to the theoretical values normalized to the H$\beta$ flux. We estimated the theoretical Balmer decrement using the data by \citet{Storey1995} assuming $T_{e}=12500$ K, and $n_{e}=100$ cm$^{-3}$. We used the extinction law of \citet{Seaton1979}. We assumed that the underlying absorption had the form $EW_{abs}({\rm H}n)=EW_{abs}({\rm H}\beta) \times g({\rm H}n)$, where $g({\rm H}n)$ was obtained from Table 2 of \citet{Pena2012}. We produced a stellar template normalized to $EW_{abs}({\rm H}\beta)$ based in the low metallicity instantaneous burst models from \citet{Gonzalez1999}, to correct for the underlying absorption equivalent widths ($EWs$) for the Balmer lines. The percent error for each line includes the uncertainties in the reddening correction and the uncertainties in underlying absorption.

The underlying absorption, reddening correction and a renormalization of H$\beta$ were done simultaneously with the equation:
\begin{eqnarray}
\label{eq:eq-int}
\frac{I(\lambda)}{I({\rm H}\beta)} & = & \frac{F(\lambda)}{F({\rm H}\beta)} 10^{f(\lambda)C({\rm H}\beta)} \left(1-\frac{EW_{abs}(\lambda)}{EW(\lambda)}\right) \nonumber \\
 & \times & \left(1-\frac{EW_{abs}({\rm H}\beta)}{EW({\rm H}\beta)}\right)^{-1}\frac{100}{\eta},
\end{eqnarray}
where $C({\rm H}\beta)$ is the reddening coefficient, $F(\lambda)$ is the absolute flux for each line, $EW(\lambda)$ is the equivalent width observed in $\lambda$ and $EW_{abs}(\lambda) = EW_{abs}({\rm H}\beta) \times g(\lambda)$ is the theoretical equivalent width \citep{Pena2012}. Only the H and He lines were corrected for underlying absorption. This underlying absorption is due to the contribution of dust scattered light, this contribution is expected to include in absorption the H and He lines present in the brightest stars.

Finally we used a $\chi^{2}$ minimization routine to obtain the parameters $C({\rm H}\beta)$, $EW_{abs}({\rm H}\beta)$, and $\eta$, the fitted values for Region I were 0.159, 1.647, and 99.53, respectively. The values for our best spectrum are in good agreement with the ones derived from \citet{Peimbert2012}. Note that H$\beta$ is not normalized to $100$ since with this value, the other lines of H, would be overestimated. In the same way that we want to average the helium lines, we want to do it with the hydrogen lines. We used the following Balmer lines H$\alpha$, H$\beta$, H$\gamma$, H$\delta$, H11, H12, and H13 to calculate the observations, this calculation implies that H$\beta=99.53$. We decided to normalize the reddening-corrected fluxes using the 7 brightest Balmer lines. The best fit to these 7 lines yielded an H$\beta$ different than 100. We took the unconventional decision of choosing this H$\beta$ value, to avoid having to correct for the optimum value of H at a later stage.

Table \ref{tab:list} shows in columns (1) and (2) the adopted laboratory wavelength $\lambda$, and the identification for each line respectively, the extinction law value used for each line \citep{Seaton1979} is in column (3), in columns (4) and (5) we present the flux $F(\lambda)$ and the intensity $I(\lambda)$ for each line (normalized to H$\beta$), and finally we show the \% error for the intensity in column (6).

\begin{longtable}{lcrrrc}
\caption{List of emission line intensities.\label{tab:list}}\\
\midrule
\midrule
\multicolumn{1}{l}{$\lambda$} &
\multicolumn{1}{c}{ID} &
\multicolumn{1}{c}{$f(\lambda)$} &
\multicolumn{1}{c}{$F(\lambda$)\tablenotemark{a}} &
\multicolumn{1}{c}{$I(\lambda$)\tablenotemark{b}} &
\multicolumn{1}{c}{\% Error} \\ \vspace{1mm}\\	\hline \vspace{1mm}\\
\endfirsthead
\caption[]{cont.}\\ \vspace{1mm}\\
\midrule
\midrule
\multicolumn{1}{l}{$\lambda$} &
\multicolumn{1}{c}{ID} &
\multicolumn{1}{c}{$f(\lambda)$} &
\multicolumn{1}{c}{$F(\lambda)$\tablenotemark{a}} &
\multicolumn{1}{c}{$I(\lambda)$\tablenotemark{b}} &
\multicolumn{1}{c}{\% Error} \\ \vspace{1mm}\\	\hline 
\endhead
3614 &\ionn{He}{I} &0.282 &0.64 &0.70 & 6 \\ \vspace{1mm}\\  
3634 &\ionn{He}{I} &0.280 &1.08 &1.18 & 4 \\ \vspace{1mm}\\  
3674 &\ionn{H}{I} &0.269 &0.40 &0.48 & 8 \\ \vspace{1mm}\\  
3679 &\ionn{H}{I} &0.268 &0.27 &0.35 & 10 \\ \vspace{1mm}\\  
3683 &\ionn{H}{I} &0.267 &0.35 &0.45 & 8 \\ \vspace{1mm}\\  
3687 &\ionn{H}{I} &0.266 &0.39 &0.52 & 8 \\ \vspace{1mm}\\  
3692 &\ionn{H}{I} &0.265 &0.55 &0.72 & 8 \\ \vspace{1mm}\\  
3697 &\ionn{H}{I} &0.264 &1.02 &1.26 & 6 \\ \vspace{1mm}\\  
3704 &\ionn{H}{I}+\ionn{He}{I} &0.262 &1.47 &1.79 & 4 \\ \vspace{1mm}\\  
3712 &\ionn{H}{I} &0.260 &1.27 &1.61 & 4 \\ \vspace{1mm}\\  
3722 &\ionn{H}{I} &0.258 &2.13 &2.32 & 4 \\ \vspace{1mm}\\  
3726 &[\ionn{O}{II}] &0.256 &30.37 &33.00 & 2 \\ \vspace{1mm}\\  
3729 &[\ionn{O}{II}] &0.255 &43.88 &47.65 & 2 \\ \vspace{1mm}\\  
3734 &\ionn{H}{I} &0.253 &2.18 &2.36 & 3 \\ \vspace{1mm}\\  
3750 &\ionn{H}{I} &0.250 &2.54 &3.14 & 3 \\ \vspace{1mm}\\  
3771 &\ionn{H}{I} &0.245 &3.33 &4.09 & 3 \\ \vspace{1mm}\\  
3789 &\ionn{C}{III}? &0.240 &0.12 &0.13 & 15 \\ \vspace{1mm}\\  
3798 &\ionn{H}{I}+[\ionn{S}{III}] &0.238 &4.14 &5.09 & 3 \\ \vspace{1mm}\\  
3820 &\ionn{He}{I} &0.233 &0.86 &1.03 & 6 \\ \vspace{1mm}\\  
3829 &\ionn{Mn}{II}? &0.231 &0.50 &0.53 & 8 \\ \vspace{1mm}\\  
3835 &\ionn{H}{I} &0.229 &6.31 &7.53 & 2 \\ \vspace{1mm}\\  
3869 &[\ionn{Ne}{III}] &0.222 &35.95 &38.58 & 1 \\ \vspace{1mm}\\  
3889 &\ionn{He}{I}+\ionn{H}{I} &0.218 &17.56 &18.82 & 1.2 \\ \vspace{1mm}\\  
3970 &\ionn{H}{I} &0.200 &25.71 &27.35 & 1 \\ \vspace{1mm}\\  
4009 &\ionn{He}{I} + [\ionn{Fe}{III}] &0.194 &0.59 &0.63 & 6 \\ \vspace{1mm}\\  
4026 &\ionn{He}{I} &0.190 &1.79 &2.07 & 4 \\ \vspace{1mm}\\  
4069 &[\ionn{S}{II}] &0.182 &1.00 &1.06 & 6 \\ \vspace{1mm}\\  
4076 &[\ionn{S}{II}] &0.181 &0.55 &0.59 & 8 \\ \vspace{1mm}\\  
4102 &\ionn{H}{I} &0.176 &24.62 &27.00 & 1 \\ \vspace{1mm}\\  
4121 &\ionn{He}{I} &0.173 &0.20 &0.30 &12 \\ \vspace{1mm}\\  
4133 &\ionn{O}{II} &0.171 &0.11 &0.11 &15 \\ \vspace{1mm}\\  
4144 &\ionn{He}{I} &0.169 &0.16 &0.27 &12 \\ \vspace{1mm}\\  
4316 &\ionn{O}{II} &0.135 &0.38 &0.39 & 8 \\ \vspace{1mm}\\  
4340 &\ionn{H}{I} &0.128 &45.95 &48.47 & 1 \\ \vspace{1mm}\\  
4363 &[\ionn{O}{III}] &0.122 &6.82 &7.06 & 2 \\ \vspace{1mm}\\  
4376 &\ionn{Ar}{II} &0.119 &0.14 &0.15 & 15 \\ \vspace{1mm}\\  
4388 &\ionn{He}{I} &0.116 &0.48 &0.56 & 8 \\ \vspace{1mm}\\  
4438 &\ionn{He}{I} &0.100 &0.23 &0.24 & 10 \\ \vspace{1mm}\\  
4472 &\ionn{He}{I} &0.094 &3.80 &4.03 & 3 \\ \vspace{1mm}\\  
4483 &\ionn{S}{II}? &0.092 &0.05 &0.05 & 20 \\ \vspace{1mm}\\  
4529 &\ionn{Ne}{I} &0.076 &0.07 &0.07 & 20 \\ \vspace{1mm}\\  
4563 &\nodata &0.070 &0.09 &0.10 & 15 \\ \vspace{1mm}\\  
4607 &[\ionn{Fe}{III}] &0.055 &0.05 &0.05 & 20 \\ \vspace{1mm}\\  
4639+42 &\ionn{O}{II} &0.051 &0.06 &0.06 & 20 \\ \vspace{1mm}\\  
4649+51 &\ionn{O}{II} &0.049 &0.12 &0.12 & 15 \\ \vspace{1mm}\\  
4658 &[\ionn{Fe}{III}] &0.047 &0.30 &0.30 & 10 \\ \vspace{1mm}\\  
4686 &\ionn{He}{II} &0.041 &0.24 &0.24 & 10 \\ \vspace{1mm}\\  
4712 &[\ionn{Ar}{IV}]+\ionn{He}{I} &0.034 &1.10 &1.10 & 6 \\ \vspace{1mm}\\  
4740 &[\ionn{Ar}{IV}] &0.028 &0.42 &0.42 & 8 \\ \vspace{1mm}\\  
4769 &[\ionn{Fe}{III}] &0.023 &0.07 &0.07 & 20 \\ \vspace{1mm}\\  
4861 &\ionn{H}{I} &0.000 &100.00 &*99.53 & 0.8 \\ \vspace{1mm}\\  
4881 &[\ionn{Fe}{III}] &-0.004 &0.16 &0.16 & 12 \\ \vspace{1mm}\\  
4922 &\ionn{He}{I} &-0.013 &1.08 &1.13 & 6 \\ \vspace{1mm}\\  
4959 &[\ionn{O}{III}] &-0.021 &178.21 &175.16 & 0.8 \\ \vspace{1mm}\\  
4986 &[\ionn{Fe}{III}] &-0.027 &0.29 &0.29 & 10 \\ \vspace{1mm}\\  
5007 &[\ionn{O}{III}] &-0.032 &534.12 &522.70 & 0.8 \\ \vspace{1mm}\\  
5016 &\ionn{He}{I} &-0.034 &2.48 &2.51 & 3 \\ \vspace{1mm}\\  
5048 &\ionn{He}{I} &-0.041 &0.13 &0.21 & 12 \\ \vspace{1mm}\\  
5192 &[\ionn{Ar}{III}]+[\ionn{N}{I}] &-0.073 &2.04 &1.97 & 4 \\ \vspace{1mm}\\  
5270 &[\ionn{Fe}{III}] &-0.089 &0.11 &0.11 & 15 \\ \vspace{1mm}\\  
5326 &\ionn{Ne}{I} &-0.096 &0.26 &0.25 & 10 \\ \vspace{1mm}\\  
5433 &[\ionn{Fe}{II}] &-0.128 &0.08 &0.08 & 20 \\ \vspace{1mm}\\  
5518 &[\ionn{Cl}{III}] &-0.140 &0.41 &0.38 & 8 \\ \vspace{1mm}\\  
5538 &[\ionn{Cl}{III}] &-0.144 &0.42 &0.39 & 8 \\ \vspace{1mm}\\  
5755 &[\ionn{N}{II}] &-0.192 &0.08 &0.07 & 20 \\ \vspace{1mm}\\  
5876 &\ionn{He}{I} &-0.216 &11.69 &10.85 & 1.5 \\ \vspace{1mm}\\  
6312 &[\ionn{S}{III}] &-0.286 &1.78 &1.61 & 4 \\ \vspace{1mm}\\  
6548 &[\ionn{N}{II}] &-0.320 &2.18 &1.95 & 4 \\ \vspace{1mm}\\  
6563 &\ionn{H}{I} &-0.322 &314.61 &282.24 & 0.8 \\ \vspace{1mm}\\  
6583 &[\ionn{N}{II}] &-0.324 &4.28 &3.83 & 3 \\ \vspace{1mm}\\  
6678 &\ionn{He}{I} &-0.337 &3.39 &3.05 & 3 \\ \vspace{1mm}\\  
6716 &[\ionn{S}{II}] &-0.342 &8.32 &7.40 & 1.5 \\ \vspace{1mm}\\  
6731 &[\ionn{S}{II}] &-0.343 &5.98 &5.31 & 2 \\ \vspace{1mm}\\  
7065 &\ionn{He}{I} &-0.383 &2.40 &2.13 & 3 \\ \vspace{1mm}\\  
7136 &[\ionn{Ar}{III}] &-0.391 &8.39 &7.33 & 2 \\ \vspace{1mm}\\  
7281 &\ionn{He}{I} &-0.406 &2.27 &1.99 & 3 \\ \vspace{1mm}\\  
7320 &[\ionn{O}{II}] &-0.410 &1.30 &1.13 & 4 \\ \vspace{1mm}\\  
7330 &[\ionn{O}{II}] &-0.411 &1.07 &0.93 & 6 \\ \vspace{1mm}\\  
\hline \vspace{1mm}\\
\multicolumn{6}{l}{\tablenotemark{a} $F(\lambda)$ in units of  $F({\rm H}\beta)=100.0$; $F({\rm H}\beta)=1.57\times10^{-12}$ erg cm$^{-2}$s$^{-1}$.} \\
\multicolumn{6}{l}{\tablenotemark{b} $I(\lambda)$ in units of  $I({\rm H}\beta)=99.53$; $I({\rm H}\beta)=2.26\times10^{-12}$ erg cm$^{-2}$s$^{-1}$.} \\
\multicolumn{6}{l}{\tablenotemark{*} Note that $I$(H$\beta)=99.53$, see text.}
\end{longtable}

To determine the [\ion{O}{2}] $\lambda\lambda3726,3729$ intensity we subtracted the contribution of the H13 and H14 Balmer lines, we also subtracted the contribution of the [\ion{S}{3}] $\lambda3722$ line that originates in the same upper level than [\ion{S}{3}] $\lambda6312$. Similarly the contribution of [\ion{Ar}{4}] to $I(4711)$ was obtained by subtracting the intensity of \ion{He}{1} $\lambda4713$. Finally the contribution of \ion{He}{1} to $I(3889)$  was obtained after subtracting the expected contribution of H8 \citep[based on the work by][]{Storey1995}.

\section{physical conditions}
\label{sec:physical}

\subsection{Temperatures and Densities from CELs}

We calculated physical conditions through CELs, fitting the line ratios using {\texttt{PyNeb}} \citep{pyneb2015}. Electron temperatures were determined using the [\ion{O}{3}] $\lambda4363/(\lambda4959+\lambda5007)$, [\ion{O}{2}] ($\lambda7320+\lambda7330$)/($\lambda3726+\lambda3729$), and [\ion{N}{2}] $\lambda5755/\lambda6584$ auroral to nebular intensity ratios. Electronic densities were determined from the [\ion{O}{2}] $\lambda3726/\lambda3729$, and [\ion{S}{2}] $\lambda6731/\lambda6716$ ratios which are strongly density dependent.

In addition, we obtained the [\ion{Fe}{3}] density from the computations by \citet{Keenan2001}, using the $I(\lambda4986)/I(\lambda4658)$ ratio. This ratio is strongly dependent on density, going from $I(4986)/I(4658)\approx1.0$ at $n_{e}=100$ cm$^{-3}$ to $I(4986)/I(4658)\approx0.05$ at $n_{e}=3000$ cm$^{-3}$. This ratio is a good density indicator, particularly for low densities, since it is more density dependent than the [\ion{Cl}{3}] and [\ion{Ar}{4}] determinations; moreover the Fe$^{++}$ density is much more representative of the whole object than the O$^{+}$ and S$^{+}$ densities, since  Fe$^{++}/$Fe$\approx60\%$ while O$^{+}/$O$\approx20\%$ and S$^{+}/$S$\approx15\%$. The temperatures and densities are presented in Table \ref{tab:tem-cel}. 

\begin{table}[htb]
\centering
\caption{Temperature and density from CELs. \label{tab:tem-cel}}
\begin{tabular}{lccc}
\hline
\hline
{Temperature (K)} & [OIII] & [OII] & [NII] \\
NGC~346 & 12871$\pm$98 & 12445$\pm$464 & 10882$\pm$767\\ \hline
Density (cm$^{-3}$) & [OII]	& [SII]	& [FeIII] \\
NGC~346 & 23.7$\pm8.2$ & 32$\pm$21	& $101\pm17$	\\ 
\hline
\end{tabular}
\end{table}

\subsection{Temperatures and Densities from RLs}
\label{subsec:tem-den_rls}

We used the program {\tt Helio14}, which is an extension of the maximum likelihood method presented by \citet{Peimbert2000} to obtain values for the temperature and density from \ion{He}{1} lines. {\tt{Helio14}} code determines the He$^{+}$/H$^{+}$ ionic abundance, $n_{e}$(\ion{He}{1}), $T_{e}$(\ion{He}{1}), the optical depth of the line \ion{He}{1} $\lambda$3889, and $t^{2}$(\ion{He}{1}) through the information on the line intensities of \ion{He}{1}, and the information of the [\ion{O}{2}] ($\lambda7320+\lambda7330$)/($\lambda3726+\lambda3729$) and the [\ion{O}{3}] $\lambda4363/(\lambda4959+\lambda5007)$ line ratios. For a full description of the method see \citet{Peimbert2007}.

By using ten \ion{He}{1} lines: $\lambda$3819, $\lambda$3889, $\lambda$4026, $\lambda$4388, $\lambda$4471, $\lambda$4922, $\lambda$5016, $\lambda$5876, $\lambda$6678, and $\lambda$7065, we obtained a temperature $T_{e}$(\ion{He}{1})$=11400\pm550$ K, and a density $n_{e}$(\ion{He}{1})$<10$ cm$^{-3}$.

\subsection{Temperature Inhomogeneities}

To derive the ionic abundance ratios, the average temperature, and the temperature standard deviation, we use the formalism of $t^{2}$ \citep{Peimbert1967}. This formalism takes into account the temperature structure inside the \ion{H}{2} region.

Each ionic species, $X^{i+}$, has a unique value of average temperature and of thermal inhomogeneity given by:
\begin{equation}
\label{eq:p01}
T_{0}(X^{i+})=\frac{\int T_{e}n_{e}n(X^{i+})dV}{\int n_{e}n(X^{i+})dV}
\end{equation}
and
\begin{equation}
\label{eq:p02}
t^{2}(X^{i+})=\frac{\int \left[T_{e}-T_{0}(X^{i+})\right]^{2}n_{e}n(X^{i+})dV}{T_{0}(X^{i+})^{2}\int n_{e}n(X^{i+})dV},
\end{equation}
where $n_{e}$ and $n(X^{i+})$ are the electron and ion densities respectively, and $V$ is the observed volume.

In general, RLs, emissivities are stronger when $T_{e}$ is lower while, CELs, emissivities are stronger when $T_{e}$ is higher; and, also in general, RL temperatures are lower than $T_0$ while CEL temperatures are higher than $T_0$. Overall, each temperature has a different $t^2$ dependence; for instance the  [\ion{O}{3}] 4363/5007 temperature, in the presence of thermal inhomogeneities, can be expressed as a function of $T_0$ and $t^2$ as:
\begin{equation}
\label{eq:p05}
T_{4363/5007} = T_{0}\left[1+\frac{t^{2}}{2}\left(\frac{91300}{T_{0}}-3\right)\right];
\end{equation}
equivalent equations can be derived for the [\ion{N}{2}], [\ion{O}{2}], and [\ion{S}{2}] temperatures as well as for \ion{He}{1} and Balmer continuum temperatures.

Since temperatures can be presented as a function of $t^{2}$ and $T_{0}$, when one measures 2 different temperatures in principle one can obtain $T_{0}$ and $t^{2}$. In practice one prefers one temperature that weighs preferably the hot regions (e.g. $T$[\ion{O}{3}]), and one that weighs preferentially the cold regions (e.g. $T$(\ion{He}{1}) or $T$(Bac)).

Combining results from CELs and values from \ion{He}{1} and using the {\tt Helio14} code we found that the maximum likelihood values are: $t^{2}$(He$^{+})=0.033\pm0.017$ and $T_0=11900\pm450$ K. These values can be used to determine abundances in the high-temperature regions, due to the similar ionization potentials of He$^{+}$ and O$^{++}$. Determinations for $t^{2}$ values from observations of \ion{H}{2} regions range between 0.020 and 0.120 \citep{Peimbert2012}. The $t^{2}$ value depends on the specific characteristics of the thermal structure of each nebula (or even of each fraction of nebula).

\section{Ionic Chemical abundances}
\label{sec:abundances}

\subsection{Heavy element ionic abundances}
\label{subsec:heavy-elements}

We have followed the $t^2$ formalism for chemical abundances as presented by \citet{Peimbert1969}. The first step to do so is to derive the abundances using the so called direct method and then to correct for the presence of thermal inhomogeneities.

We have used a two-zone direct method characterized by the zones where low- and high-ionization ions are present: $T$(low) and $T$(high). $T$(low) is the mean of $T_e$[\ion{N}{2}], $T_e$[\ion{O}{2}] and $T_e$[\ion{S}{2}], while $T$(high) is $T_e$[\ion{O}{3}]; we used $T$(low) for singly ionized heavy elements, and $T$(high) for multiple ionized elements.

We used {\texttt{PyNeb}} to made the computations \citep{pyneb2015}. In Table {\ref{tab:abion}} we present two values for each ionic abundance, the first value is obtained directly from {\texttt{PyNeb}} ($t^{2}=0.00$), and the second value is using the formalism of $t^{2}$.

To obtain the ionic abundance using $t^{2}\neq0.00$, we used the equations by \citet{Peimbert2004}:

\begin{eqnarray}
\label{ab-t2}
\left[\frac{n_{\it cel}(X^{+i})}{n({\rm H}^{+})}\right]_{t^{2}\neq0.00}& = & \frac{T({\rm H}\beta)^{\alpha}T(\lambda_{nm})^{0.5}}{T_{(4363/5007)}^{\alpha+0.5}} \nonumber \\
 & \times & \exp\left[-\frac{\Delta E_{n}}{kT_{(4363/5007)}}+\frac{\Delta E_{n}}{kT(\lambda_{nm})}\right] \nonumber \\
 & \times & \left[\frac{n_{\it cel}(X^{+i})}{n({\rm H}^{+})}\right]_{t^{2}=0.00},
\end{eqnarray}
where $\alpha=-0.89$ is the termperature dependence of H$\beta$, $\Delta E_{n}$ is the difference of energy between the ground and the excited level of the CEL, and $T(\lambda_{nm})$ and $T({\rm H}\beta)$ are line temperatures, as described by \citet{Peimbert1967} (and represent the average emmisivity of each line for the temperature distribution on the observed object) and can be written as:
\begin{equation}
\label{eq:p03}
T(\lambda_{nm})=
T_{0}\left\lbrace 1+\frac{t^{2}}{2}\left[\frac{\left(\Delta E_{n}/kT_{0}\right)^{2}-3\Delta E_{n}/kT_{0}+3/4}{\Delta E_{n}/kT_{0}-1/2}\right]
\right\rbrace, 
\end{equation}
and
\begin{equation}
\label{eq:p04}
T({\rm H}\beta)=T_{0}\left[1-\frac{t^{2}}{2}(1-\alpha)\right].
\end{equation}

\begin{table}[htb]
\centering
\caption{Ionic abundances from collisional excitation lines for NGC~346. \label{tab:abion}}
\begin{tabular}{lcc}
\hline
\hline
Ion  & $t^{2}=0.000$  & $t^{2}=0.033\pm0.017$ \\ 
\hline
N$^{+}$		& 5.87$\pm$0.02	& 5.97$\pm$0.06	\\
O$^{+}$		& 7.41$\pm$0.05	& 7.54$\pm$0.08	\\
O$^{++}$	& 7.93$\pm$0.01	& 8.04$\pm$0.06	\\
Ne$^{++}$	& 7.22$\pm$0.01	& 7.33$\pm$0.06	\\
S$^{+}$		& 5.55$\pm$0.03	& 5.65$\pm$0.06	\\
S$^{++}$	& 6.18$\pm$0.02	& 6.24$\pm$0.04	\\
Ar$^{++}$	& 5.54$\pm$0.01	& 5.60$\pm$0.03	\\
Ar$^{+3}$	& 5.11$\pm$0.02	& 5.22$\pm$0.06	\\
Cl$^{++}$	& 4.37$\pm$0.03	& 4.47$\pm$0.06	\\
\hline
\multicolumn{3}{c}{In units of $12+\log n(X^{+i})/n({\rm H})$.}
\end{tabular}
\end{table}

\subsection{Helium Abundance}
\label{subsec:abun-He}

To determine the He abundance we also used the 10 lines mentioned in Sec. \ref{subsec:tem-den_rls} as input for the {\tt Helio14} code. This code uses the effective recombination coefficients given by \citet{Storey1995} for  H$^{+}$ and for the case of He$^{+}$ those given by \citet{Benjamin2002} and \citet{Porter2013}. The collisional contribution was estimated from \citet{Sawey1993} and \citet{Kingdon1995}, and we used the calculations made by \citet{Benjamin2002} for the optical depth effects in the triplets. 

The correction of the underlying stellar absorption is important for \ion{He}{1} lines, therefore we used the values from \citet{Gonzalez1999} to correct lines with $\lambda<5000$\AA, and for lines redder than 5876\AA, we used the values of \citet{Peimbert2005}. The ionic abundance of He$^{++}$ was obtained from the $\lambda4686$ line presented in Table \ref{tab:list}, and the recombination coefficients given by \citet{Storey1995}.

\begin{table}[htb]
\centering
\caption{Ionic abundances from recombination lines for NGC~346. \label{tab:abHe}}
\begin{tabular}{lcc}
\hline
\hline
Ion			& $t^{2}=0.000$ & $t^{2}=0.033\pm0.017$ \\ \hline
He$^{+}$	& 10.917$\pm$0.004	& 10.915$\pm$0.004	\\
He$^{++}$	& 8.30$\pm$0.04		& 8.30$\pm$0.04		\\
\hline
\multicolumn{3}{c}{In units of $12+\log n(X^{+i})/n({\rm H})$.}
\end{tabular}
\end{table}

\section{Total abundances}
\label{sec:total-abun}

For most elements we cannot observe all their ionic stages; so, in general, ionization correction factors (ICFs) are required. The functional forms of those ICFs frequently depend on the depth, quality, and coverage of the spectra.  

In general, for helium:
\begin{eqnarray}
\label{eq:he}
\frac{N({\rm He})}{N({\rm H})} & = & \frac{N({\rm He}^{0})+N({\rm He}^{+})+N({\rm He}^{++})}{N({\rm H}^{0})+N({\rm H}^{+})} \nonumber \\
 & = & {\rm ICF}({\rm He}^{0})
\frac{N({\rm He}^{+})+N({\rm He}^{++})}{N({\rm H}^{+})},
\end{eqnarray}
for objects of low, and medium, degree of ionization, the presence of neutral helium within the \ion{H}{2} region is important and the ${\rm ICF(He^0)}>1$; for \ion{H}{2} regions with a high degree of ionization, as is NGC~346, we can consider that the presence of neutral helium within the nebula is negligible; however one must be careful as for objects with very high degree of ionization the ICF(He$^0$) < 1.00 (even if the correction for helium is always ${\rm ICF(He^0)} \gtrsim 0.99$, when we are looking for a precise determination of $Y_{\rm P}$, one should try to avoid such objects).

To determine an accurate He/H ratio a precise value of the ICF(He$^0$) is needed. Previous computations by \citet{Luridiana2003,Peimbert2002,Peimbert2007,Relano2002} showed that the ICF(He$^0$)$=1.00$ for NGC~346, because there is no transition zone from ionized to neutral He and H regions. We note that their results were calculated using tailor-made photoionization models  with the code {\texttt{CLOUDY}} \citep{Ferland2013}.

For oxygen, to obtain the total abundance, we considered that:
\begin{equation}
\frac{N({\rm O}^{3+})}{N({\rm O})} = \frac{N({\rm He}^{++})}{N({\rm He})},
\end{equation}
and
\begin{equation}
{\rm ICF}({\rm O}^{+}+{\rm O}^{++}) = \frac{N({\rm O}^{+})+N({\rm O}^{++})+N({\rm O}^{3+})}{N({\rm O}^{+})+N({\rm O}^{++})};
\end{equation}
therefore:
\begin{equation}
\frac{N({\rm O})}{N({\rm H})} = {\rm ICF}({\rm O}^{+}+{\rm O}^{++})\left[\frac{N({\rm O}^{+})+N({\rm O}^{++})}{N({\rm H}^{+})}\right],
\label{eq:oxy}
\end{equation}
where
\begin{equation}
{\rm ICF}({\rm O}^{+}+{\rm O}^{++}) = \left[1-\frac{N({\rm He}^{++})}{N({\rm He})}\right]^{-1}.
\end{equation}

For nitrogen, and neon we obtained the total abundances using the ICFs by \citet{Peimbert1969}.
\begin{equation}
\frac{N({\rm N})}{N({\rm H})} = {\rm ICF}({\rm N}^{+})\left[\frac{N({\rm N}^{+})}{N({\rm H}^{+})}\right],
\label{eq:n}
\end{equation}
where 
\begin{equation}
{\rm ICF}({\rm N}^{+}) = \frac{N({\rm O}^{+})+N({\rm O}^{++})+N({\rm O}^{+3})}{N({\rm O}^{+})},
\end{equation}
and
\begin{equation}
\frac{N({\rm Ne})}{N({\rm H})} = {\rm ICF}({\rm Ne}^{++})\left[\frac{N({\rm Ne}^{++})}{N({\rm H}^{+})}\right],
\label{eq:ne}
\end{equation}
where 
\begin{equation}
{\rm ICF}({\rm Ne}^{++}) = \frac{N({\rm O}^{+})+N({\rm O}^{++})+N({\rm O}^{+3})}{N({\rm O}^{++})}.
\end{equation}

We observed the auroral lines of [\ion{S}{2}] and [\ion{S}{3}]. To obtain the total sulphur abundance we used the ICF proposed by \citet{Stasinska1978}:
\begin{equation}
\frac{N({\rm S})}{N({\rm H})} = {\rm ICF}({\rm S}^{+}+{\rm S}^{++})\left[\frac{N({\rm S}^{+})+N({\rm S}^{++})}{N({\rm H^{+}})}\right]
\end{equation}
where
\begin{equation}
{\rm ICF}({\rm S}^{+}+{\rm S}^{++}) = \left[1-\left(1-
\frac{N({\rm O}^{+})}{N({\rm O})}\right)^{3}\right]^{1/3}.
\end{equation}

For the case of argon, we have measurements for lines of [\ion{Ar}{3}] and [\ion{Ar}{4}]; we used the calculations by \citet{PerezM2007}:
\begin{equation}
\frac{N({\rm Ar})}{N({\rm H})} = {\rm ICF}({\rm Ar}^{++}+{\rm Ar}^{3+})
\left[\frac{N({\rm Ar}^{++})+N({\rm Ar}^{+3})}{N({\rm H}^{+})}\right]
\label{eq:ar}
\end{equation}
where
\begin{equation}
{\rm ICF}({\rm Ar}^{++}+{\rm Ar}^{3+}) = 0.928+0.364\left(1-x\right)+\frac{0.006}{1-x}.
\end{equation}

For chlorine, we have measurements for lines of [\ion{Cl}{3}], and to obtain the total abundance we used a reimplementation of the ICF proposed by \citet{Delgado2014}:
\begin{equation}
\frac{N({\rm Cl})}{N({\rm H})} = {\rm ICF}({\rm Cl}^{++})\frac{N({\rm O}^{+})+N({\rm O}^{++})}{N({\rm H}^{+})}
\frac{N({\rm Cl}^{++})}{N({\rm O}^{+})},
\end{equation}
where
\begin{equation}
{\rm ICF}({\rm Cl}^{++}) = 2.914 \left(1-\left[ \frac{N({\rm O}^{++})}{N({\rm O}^{+})+{N({\rm O}}^{++})} \right]^{0.21}\right)^{0.75}.     
\end{equation}

In Table \ref{tab:tot-cel} we present the results for $t^{2}=0.00$ as well as for $t^{2}=0.033\pm0.017$.

\begin{table}[htb]
\centering
\caption{Total abundances for NGC~346. \label{tab:tot-cel}} 
\begin{tabular}{ccc}
\hline
\hline 
Element  & $t^{2}=0.000$  & $t^{2}=0.033\pm0.017$ \\ 
\hline
He	& 10.918$\pm$0.004  & 10.916$\pm$0.004 	\\
N	& 6.50$\pm$0.03	&	6.61$\pm$0.07 \\
O	& 8.05$\pm$0.02	&	8.19$\pm$0.08 \\
Ne	& 7.33$\pm$0.02	&	7.48$\pm$0.08 \\
S	& 6.40$\pm$0.03	&	6.44$\pm$0.08 \\
Ar	& 5.70$\pm$0.02	&	5.82$\pm$0.07 \\
Cl	& 5.37$\pm$0.04	&	5.47$\pm$0.07 \\
\hline
\multicolumn{3}{c}{In units of $12+\log n(X)/n({\rm H})$.}
\end{tabular}
\end{table}

\subsection{$X$, $Y$, and $Z$ values for NGC~346}
\label{subsec:xyz}

To calculate the fraction of He by mass, we need to know the fraction of the heavy elements by mass. For low metallicity objects, like NGC~346, the $O/Z$ ratio is expected to be $O/Z\approx0.55$ \citep{Peimbert2007}; this ratio is slightly larger than for high metallicity objects where $O/Z\sim0.40$, mainly due to the increase of the $C/O$ ratio \citep{Carigi2011}. Using $O/Z=0.55$ thus we can estimate the full value of $Z=0.0030$ and from that $Y=0.2490$.

One of the most subtle effects that we need to include in the determination of primordial helium or helium in very hot objects, is the correction due to the collisional excitation of the Balmer lines. Since models in \citet{Peimbert2002} that were later used in \citet{Peimbert2007,Peimbert2016}, we know that for NGC~346 to include this effect produces an increase of $0.0015\pm0.0005$, consequently $X=0.7465$, $Y=0.2505$ and $Z=0.0030$. In this work, the contribution by collisions to the intensity of He and H lines is considerably smaller than objects with higher electron temperature like SBS~$0335-052$ or I~Zw~18. To minimize this particular error it is important to study $O/H$ poor objects but not extremely poor. Therefore, if we wish to further minimize this error, we must observe colder objects, with heavy elements abundances similar to those of NGC~346.

\section{Primordial helium abundance}
\label{sec:prim-helium}

To obtain the $Y_{\rm P}$ value, we need to compute the fraction of helium present in the interstellar medium produced by galactic chemical evolution. For this purpose it was assumed that:
\begin{equation}
Y_{\rm P}=Y-O\phantom{1}\frac{\Delta Y}{\Delta O},
\end{equation}
where $Y$ and $O$ are the helium and oxygen abundances by mass. We adopted the determination  $\Delta Y / \Delta O = 3.3\pm0.7$ by \citet{Peimbert2016}. To obtain this value, they used the observations of brighter objects from \citet{Peimbert2007} and chemical the evolution models for galaxies of low mass and metallicity by \citet{Carigi2008} and \citet{PeimbertIAU2010}.

Most determinations are dominated by systematic errors rather than statistical errors. We can decrease the statistical errors increasing the number of objects used to determine $Y_{\rm P}$. Nevertheless, the systematic ones will not decrease.

\begin{table}[htb]
\centering
\caption{Error budget in the $Y_{\rm P}$ determination. \label{tab:error}} 
\begin{tabular}{lccc}
\hline
\hline
Problem & Error & * &  This paper \\ 
\hline
$\Delta O(\Delta Y/ \Delta O)$ correction   & Systematic    & $\pm0.0010$ &  $\pm0.0012$  \\
Rec. coefficients of \ion{He}{1}            & Systematic    & $\pm0.0010$ &  $\pm0.0010$  \\
Temperature structure                       & Statistical   & $\pm0.0010$ &  $\pm0.0010$  \\
Reddening correction                        & Systematic    & $\pm0.0007$ &  $\pm0.0007$  \\
Collisional excitation of \ion{He}{1}       & Statistical   & $\pm0.0007$ &  $\pm0.0007$  \\
Underlying absorption in \ion{He}{1}        & Statistical   & $\pm0.0007$ &  $\pm0.0007$  \\
Underlying absorption in \ion{H}{1}         & Statistical   & $\pm0.0005$ &  $\pm0.0007$  \\
Optical depth of \ion{He}{1} triplets       & Statistical   & $\pm0.0005$ &  $\pm0.0007$  \\
Collisional excitation of \ion{H}{1}        & Systematic    & $\pm0.0015$ &  $\pm0.0005$  \\
Rec. coefficients of \ion{H}{1}             & Systematic    & $\pm0.0005$ &  $\pm0.0005$  \\
\ion{He}{1} and \ion{H}{1} line intensities & Statistical   & $\pm0.0005$ &  $\pm0.0005$  \\
He ionization correction factor             & Statistical   & $\pm0.0005$ &  $\pm0.0003$  \\
Density structure                           & Statistical   & $\pm0.0005$ &  $\pm0.0003$  \\
\hline
\multicolumn{3}{l}{*\citet{Peimbert2007}.} \\
\end{tabular}
\end{table}

In Table \ref{tab:error}, we present the error budget of our determination. The error is derived from the same 13 sources considered in \citet{Peimbert2007}. The most important source of error is due to the difficulty to correct for the collisional excitation of the Balmer lines; in very hot objects, $T_e \ge 15000$, this excitation produces a non negligible increase in the Balmer line intensities relative to the case B recombination and, if ignored, introduces a bias in the reddening correction deduced from the Balmer decrement, both effects affecting the calibration of all the lines in the spectra by up to a few percent; however, since this effect has not been studied extensively the corrections that can be applied are uncertain at best, and thus such objects should be avoided. Another important source of error comes from the temperature structure; this arises because most Y determinations are based on a single homogeneous temperature, $T(4363/5007)$, however the $T_e($\ion{He}{1}$)$ is systematically lower showing the presence of temperature variations that should be included in the $Y_{\rm P}$ determination. The error of $O(\Delta Y/\Delta O)$ correction implies the extrapolation to zero heavy-element content; based on chemical evolution models of galaxies of different types, it is found that $Y/O$ is practically constant for objects with $O<4\times10^{-3}$. To estimate the error in the reddening correction, we made comparisons among four classic extinction laws and a recent one; these laws are labeled S79 \citep{Seaton1979}, W58 \citep{Whitford58}, CCM89 \citep{Cardelli89}, and B07 \citep{Blagrave07}. The systematic effect of $ICF$(He), has to be tested with tailor-made photoionization models for low-metallicity \ion{H}{2} regions. A discussion of the remaining (smaller) sources errors is presented in \citet{Peimbert2007}.

When compared with errors of the sample of 2007 we can see that some systematic errors go down because we are using the best-known object to determine $Y_{\rm P}$; while a few of the statistical errors go up because we are using a single object with a notable increase in $\Delta Y/ \Delta O$. One of the advantages of NCG~346 is that the correction due to the collisional excitation of the Balmer lines is smaller $0.0015\pm0.0005$ in comparison with $0.0144\pm0.0038$ for SBS $0335-052$, and the $0.0056\pm0.0015$ for the sample by \citet{Peimbert2007,Peimbert2016}. Overall the total error goes down, and the $Y_{\rm P}$ value amounts to $0.2451\pm0.0026$.

We can also compare with previous determinations for $Y_{\rm P}$ determined using only NCG~346: $Y_{\rm P}=0.2345\pm0.0026$ by \citet{Peimbert2000}, $Y_{\rm P}=0.2384\pm0.0025$ by \citet{Peimbert2002}, $Y_{\rm P}=0.2453\pm0.0033$ by \citet{Peimbert2007}, and $Y_{\rm P}=0.2433\pm0.0034$ by \citet{Peimbert2016}. The difference when comparing the values of 2000 and 2002 with those from 2007 and onward comes from a better model of NGC346, newer atomic data and a better understanding of the quantity and magnitude of the sources of error (particularly, the effects of the collisional excitation of the Balmer lines were ignored in the first two determinations).

To summarize there are several advantages in observing a nearby \ion{H}{2} region. The higher spatial resolution allows us to isolate the windows that are contaminated by the light of bright stars. This allows us to reduce the starlight emission and therefore to obtain a greater signal to noise in the emission lines. In this determination, the systematic errors due to collisional excitation of the H are smaller and some statistical errors (like the ICF(He$^0$) and the density structure), also become smaller.

\begin{figure}[htb]
\begin{center}
\includegraphics[width=9.0cm]{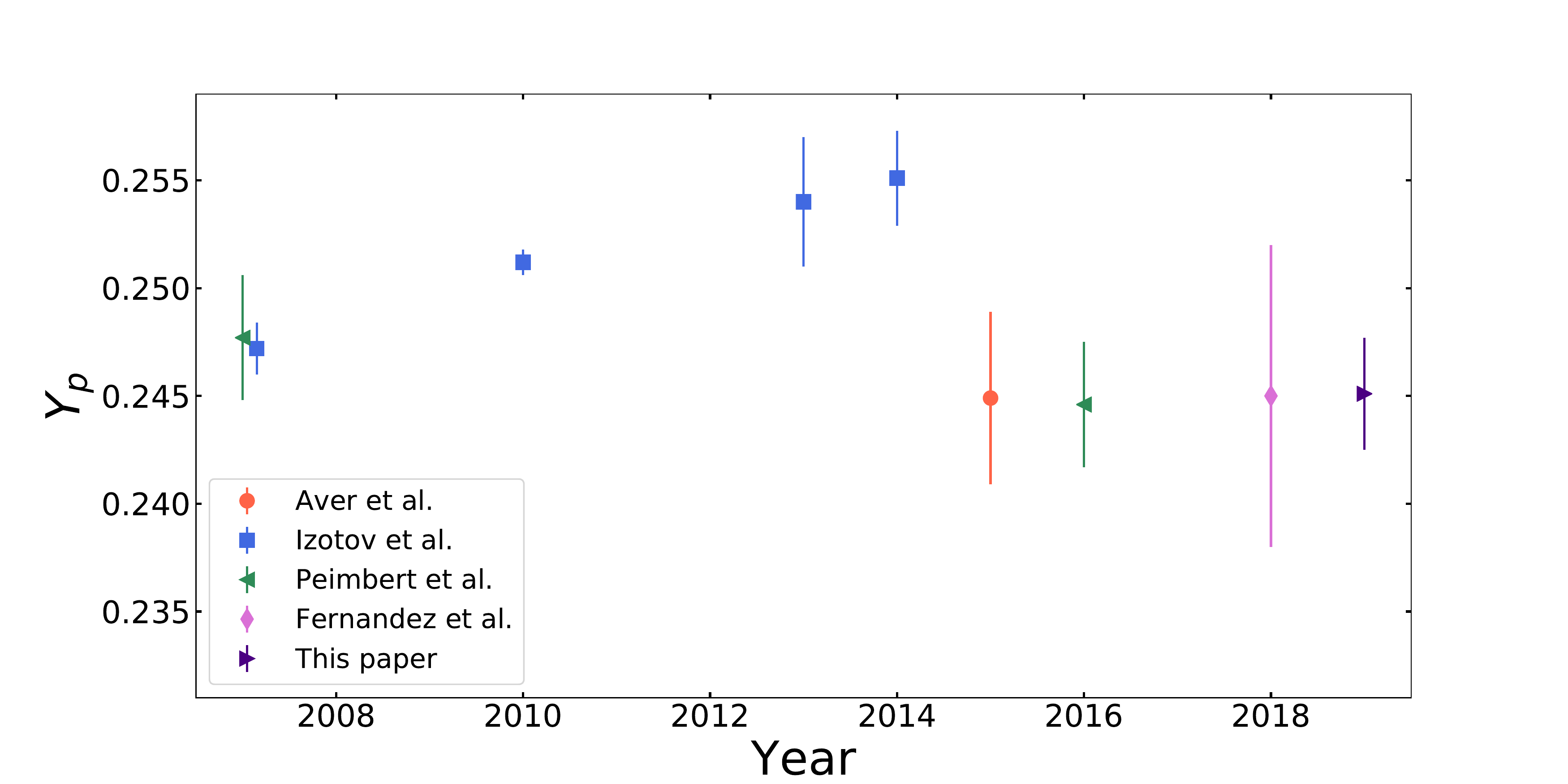}
\caption{Measurements $Y_{\rm P}$ of the last 12 years, using \ion{H}{2} regions.}
\label{fig:history}
\end{center}
\end{figure}

It is useful to discuss the determination of $Y_{\rm P}$ as a function time, \citet{Skillman2012} presents the plot with results from 1974 to 2012, accordingly in Figure \ref{fig:history} we present the $Y_{\rm P}$ results from 2007 to the present. It is clear that the $Y_{\rm P}$ values derived by different groups provide an important constraint to the big bang theory. The main source of error is due to systematic uncertainties. In addition, the result by the \citet{Plank2016}, corresponds to $Y_{\rm P}=0.24467\pm0.0002$.

\section{Discussion and conclusions}
\label{sec:conclusions}

Low-metallicity \ion{H}{2} regions are optimal objects to determine the primordial helium abundance. In this context, we studied the \ion{H}{2} region NGC~346. By measuring its \ion{He}{1} line intensities, we determined a primordial helium abundance $Y_{\rm P}=0.2451\pm0.0026$.

Table \ref{tab:neutrino} shows $Y_{\rm P}$ measurements reported in the literature and in this work. Our $Y_{\rm P}$ value is consistent with most of the previous estimates. We note that the \citet{Izotov2014} determination differs significantly from the other values. We consider that they have some systematic errors that they are ignoring; for example, part of this difference is due to the adopted temperature structure: \citet{Izotov2014} used the temperature from \texttt{CLOUDY} photoionized models \citep{Ferland2013}, which predict $0.000<t^2<0.015$, with a typical value of about $0.004$ \citep{Peimbert-Gloria2017}; alternatively the observed $t^2$ values for 28 \ion{H}{2} regions are in the $0.019$ to $0.120$ range with an average value of $0.044$ \citep{Peimbert2012}. A discussion on the possible reasons for the high $t^2$ values observed in \ion{H}{2} regions is given by \citet{Peimbert2016}.

Since the regions observed in all cases are not the same, we do not expect the values to be the same, e.g. there may be regions where the $t^2$ value may be higher due to stellar winds or other sources of energy input \citep[e.g.,][]{Peimbert-Gloria2017}.

According to our error budget presented in Table \ref{tab:error} it is comforting that three of the other four independent $Y_{\rm P}$ results are very similar to ours. The main points of our result are: a) the lower temperature of NGC~346 reduces considerably the collisional excitation of the hydrogen lines as well as the uncertainty of its determination, when compared to the other samples, all of which include a few very hot objects, and b) the small error introduced by correcting for the underlying absorption of the \ion{He}{1} lines in NGC~346, this is because we were able to eliminate most of the contribution of the dust scattered stellar light in our observations.

\citet{Peimbert2016} presents a $Y_{\rm P}$ value determined only using NGC~346; that determination has its error divided into statistical and systematical components that amounts $0.0028$ and $0.0019$ respectively, while in this paper the statistical error has been diminished to $0.0018$ and the systematical one remains at $0.0019$. The systematical error was not expected to change, since we are using the same atomic data, the same reddening law, etc.; on the other hand the statistical error was indeed expected to diminish due to the greater quality of the data (the bigger telescope and the higher number of photons collected).

While this is a significant improvement over previous determinations, we would still like better determinations. To continue the quest to obtain a $Y_{\rm P}$ value of higher precision, we need: a) helium atomic data of higher quality; b) better chemical evolution models of dwarf galaxies, that would allow us to have a better determination of the $\Delta Y/\Delta O$ ratio; and c) a newer study of the reddening law, one designed to take full advantage of the newer telescopes and instruments that were not available a few decades ago. We also consider that observations, of objects like NCG~346,  with the future generation extra-large telescopes  will help reduce the error in the $Y_{\rm P}$ determination.

\begin{table}[htb]
\centering
\caption{$Y_{\rm P}$ values and predicted equivalent number of neutrino families, $N_{\nu}$. \label{tab:neutrino}} 
\begin{tabular}{lcc}
\hline
\hline
$Y_{\rm P}$ source & $Y_{\rm P}$ & $N_{\nu}$  \\ 
\hline 
\citet{Izotov2014}    &  $0.2551\pm0.0022$ & $3.58\pm0.16$ \\
\citet{Aver2015}      &  $0.2449\pm0.0040$ & $2.91\pm0.30$ \\
\citet{Peimbert2016}  &  $0.2446\pm0.0029$ & $2.89\pm0.22$ \\
\citet{Vital2018}     &  $0.245\pm0.007$   & $2.92\pm0.55$ \\
This work             &  $0.2451\pm0.0026$ & $2.92\pm0.20$ \\
\hline
\multicolumn{3}{l}{Assuming $\tau_{n}= 880.2$ s \citep{Data2018}.}
\end{tabular}
\end{table}

We estimated the number of neutrino families, $N_{\nu}$, using the $Y_{\rm P}$ values reported in Table \ref{tab:neutrino} assuming a neutron half-life value $\tau_n=880.2\pm1.0$s by \citet{Data2018}. Table \ref{tab:neutrino} shows the $N_{\nu}$ values from the previous cited authors. As expected the result in this paper is consistent with the results by \citet{Aver2015}, \citet{Peimbert2016}, and \citet{Vital2018}; these results agree with the presence of three families of neutrinos, which is consistent with laboratory determinations \citep[e.g.,][]{Mangano2005, Mangano2011}. While the $Y_{\rm P}$ result by \citet{Izotov2014} suggests the presence of a fourth neutrino family. which would not be fully ultrarelativistic (light) at the time of neutrino decoupling.

\begin{table}[htb]
\centering
\caption{$Y_{\rm P}$ values and the neutron mean life, $\tau_{n}$. \label{tab:tau}} 
\begin{tabular}{lcc}
\hline
\hline
$Y_{\rm P}$ source & $Y_{\rm P}$ & $\tau_{n}$  \\ 
\hline 
\citet{Izotov2014}    &  $0.2551\pm0.0022$ & $921\pm11$ \\
\citet{Aver2015}      &  $0.2449\pm0.0040$ & $872\pm19$ \\
\citet{Peimbert2016}  &  $0.2446\pm0.0029$ & $870\pm14$ \\
\citet{Vital2018}     &  $0.245\pm0.007$   & $872\pm33$ \\
This work             &  $0.2451\pm0.0026$ & $873\pm13$ \\
\hline
\multicolumn{3}{l}{Assuming $N_{\nu}= 3.046$ \citep{Mangano2005}.}
\end{tabular}
\end{table}

We also estimate the value of the neutron mean life, $\tau_{n}$ using the value of $N_{\nu}$ reported by \citet{Mangano2005}, together with the $Y_{\rm P}$ estimations in Table \ref{tab:tau}. The results obtained from \citet{Aver2015}, \citet{Peimbert2016}, \citet{Vital2018}, and ourselves, are within 1$\sigma$ from the value presented by \citet{Data2018}, and they are in agreement with the $\tau_{n}$ laboratory determinations. Alternatively from the $Y_{\rm P}$ results obtained by \citet{Izotov2014}, the $\tau_{n}$ value differs by more than 3$\sigma$ from the laboratory determination.

\acknowledgments
We are grateful to the anonymous referee for a careful reading of the manuscript and some excellent suggestions. We are indebted to Maria Teresa Ruiz for her collaboration in the observations of this object. We acknowledge CONACyT for supporting this project throught grant 241732. M. V. acknowledges several constructive suggestions in Python by Oscar Barrag\'an.

\bibliographystyle{yahapj}
\bibliography{references_rev}

\end{document}